\documentclass[11pt]{article}
\usepackage{amsmath}
\usepackage{amssymb}

\usepackage[english]{babel}
\usepackage{latexsym}
\usepackage{amstext, amsfonts,amsbsy}

\usepackage{colortbl}
\usepackage{longtable}

\catcode`\@=11
\def\marginnote#1{}

\newcommand{\tr}{{\rm tr}}
\newcommand{\ti}[1]{\tilde{#1}}

\newcommand{\la}{\lambda}
\newcommand{\vf}{\varphi}
\newcommand{\al}{\alpha}
\newcommand{\be}{\beta}

\newcommand{\om}{\omega}
\newcommand{\vth}{\vartheta}

\newcommand{\Mat}{ {\rm Mat}(N,\mathbb C) }

\newcommand{\mC}{\mathbb C}
\newcommand{\mZ}{\mathbb Z}

\def\beq{\begin{equation}}
\def\eq{\end{equation}}
\def\p{\partial}

\def\res{\mathop{\hbox{Res}}\limits}

\begin{document}

\setcounter{page}{1}

\

\vspace{-15mm}

\begin{flushright}
% ITEP-TH-??/21\\
\end{flushright}
\vspace{5mm}

%\date{}
%\date{}
%\vspace{50mm}
\begin{center}
\vspace{-8mm}
{\Large{\bf Higher rank 1+1 integrable Landau-Lifshitz field theories }
 \\ \vspace{4mm}
{\Large{\bf from associative Yang-Baxter equation}}
%{\LARGE{on ${ GL}(NM)$-bundles over elliptic curve}
}

 \vspace{16mm}

 {\Large  {K. Atalikov}\,\footnote{Steklov Mathematical Institute of Russian
Academy of Sciences, Gubkina str. 8, 119991, Moscow, Russia;
 %Moscow
 %Institute of Physics and Technology, Inststitutskii per.  9,
 %Dolgoprudny, Moscow region, 141700, Russia;
 Institute for Theoretical and Experimental Physics of NRC ''Kurchatov Institute'',
 B.Cheremushkinskaya 25, Moscow 117218, Russia;
 e-mail:
 kantemir.atalikov@yandex.ru.}
 \quad\quad\quad
 {A. Zotov}\,\footnote{Steklov Mathematical Institute of Russian
Academy of Sciences, Gubkina str. 8, 119991, Moscow, Russia;
National Research University Higher School of Economics,
Usacheva str. 6,  Moscow, 119048, Russia;
 %Institute for Theoretical and Experimental Physics of NRC ''Kurchatov Institute'',
 %B.Cheremushkinskaya 25, Moscow 117218, Russia;
 % National Research
 % University Higher School of Economics, Russian Federation;
%  Moscow
% Institute of Physics and Technology, Inststitutskii per.  9,
% Dolgoprudny, Moscow region, 141700, Russia;
 e-mail: zotov@mi-ras.ru.}
 }
\end{center}

\vspace{5mm}

\begin{abstract}
 We propose a construction of 1+1 integrable Heisenberg-Landau-Lifshitz type equations in the ${\rm gl}_N$ case.
 The dynamical variables are matrix elements of $N\times N$ matrix $S$ with the property $S^2={\rm const}\cdot S$. The Lax pair with spectral parameter is constructed by means of a quantum $R$-matrix satisfying the associative Yang-Baxter equation. Equations of motion for ${\rm gl}_N$ Landau-Lifshitz model are derived from the Zakharov-Shabat equations. The model is simplified when ${\rm rank}(S)=1$. In this case the Hamiltonian description is suggested. The described family of models includes the elliptic model coming from ${\rm GL}_N$ Baxter-Belavin elliptic $R$-matrix. In $N=2$ case the widely known Sklyanin's elliptic Lax pair for XYZ Landau-Lifshitz equation is reproduced. Our construction is also valid for trigonometric and rational degenerations of the elliptic $R$-matrix.
 %In this way we obtain higher rank generalizations of XXZ and XXX classical Heisenberg magnets.
 %including their 7-vertex and 11-vertex deformations.
\end{abstract}

%\bigskip

\bigskip

%\newpage

%\small{
%\tableofcontents
%}
%\newpage
\parskip 5pt plus 1pt   \jot = 1.5ex

%%%%%%%%%%%%%%%%%%%%%%%%%%%%%%%%%%%%%%%%%%%%%%%%%%%%%%%%%%%%%%%%%%%%%%%%%%%%%%%%%%%%%%%%%%%%%%%%%%%%%%
%%%%%%%%%%%%%%%%%%%%%%%%%%%%%%%%%%%%%%%%%%%%%%%%%%%%%%%%%%%%%%%%%%%%%%%%%%%%%%%%%%%%%%%%%%%%%%%%%%%%%%

%\subsection*{Introduction: brief review and summary}
\subsection*{\underline{Introduction and notations}}
\paragraph{The Landau-Lifshitz equation \cite{LL}} is 1+1 field theory describing behaviour of  the magnetization vector $\vec{S}(t,x)=(S_1,S_2,S_3)$
  in one-dimensional model of ferromagnetic solid ($x$ is a space variable on a unit circle and $t$ is a time variable):
  \beq\label{q01}
  \begin{array}{l}
  \displaystyle{
 \p_t {\vec S}={\ti c}_1 {\vec S}\times J({\vec S})+{\ti c}_2 {\vec S}\times \p_x^2{\vec S}\,,
 \qquad J({\vec S})=(J_1S_1,J_2S_2,J_3S_3)\,,
 }
 \end{array}
 \eq
where %$''\times''$ is the vector product,
${\ti c}_1\,,{\ti c}_2$ and $J_1\,,J_2\,,J_3$ are some constants. The periodic boundary conditions
${\vec S}(t,x)={\vec S}(t,x+2\pi)$ are assumed. Also, for simplicity  we deal with the complex version of (\ref{q01}) so that
the dynamical variables $(S_1,S_2,S_3)$ and all the constants are $\mC$-valued (the reduction to real-valued case is available as well). The second term in the r.h.s. of (\ref{q01}) describes the spin exchange interaction, while the first one comes from the anisotropy. In the fully anisotropic case (when $J_1\,,J_2\,,J_3$ are pairwise distinct)
the model is referred to as XYZ type model, the partially anisotropic case ($J_1=J_2$) is called XXZ, and the fully isotropic case ($J_1=J_2=J_3$) is known as XXX model. In the latter case ${\vec S}\times J({\vec S})=0$ and (\ref{q01})
is reduced to the equation for 1+1 continuous classical XXX Heisenberg magnet.
Equivalently, the Landau-Lifshitz equation (\ref{q01}) is written in the matrix form
  \beq\label{q02}
  \begin{array}{l}
  \displaystyle{
 \p_t {S}={c}_1 [{S}, J({S})]+{c}_2 [{S}, \p_x^2{S}]\,,
 \qquad S=\sum\limits_{k=1}^3S_k\sigma_k\,,\quad J(S)=\sum\limits_{k=1}^3S_k J_k\sigma_k\,,
 }
 \end{array}
 \eq
 where $S$ is a traceless $2\times 2$ matrix, $S_k$ -- its components in the basis of the Pauli matrices $\sigma_k$ and $c_{1,2}=2\sqrt{-1}{\ti c}_{1,2}$.

\paragraph{Lax pair in $2\times 2$ case.} Integrability of (\ref{q01})-(\ref{q02}) was established in \cite{Skl} (see also \cite{FT}) in the sense of the classical inverse scattering method, which is based on representing a nonlinear equation(s) in the Zakharov-Shabat (or zero-curvature) form \cite{ZaSh,FT}:
  \beq\label{q03}
  \begin{array}{l}
  \displaystyle{
 \p_t U(z)-\p_x V(z)+[U(z),V(z)]=0\,,
 }
 \end{array}
 \eq
 where $U(z),V(z)$ is a pair of matrix-valued functions of the variables $t,x$ depending also on the spectral parameter $z$. In \cite{Skl} the following Lax pair (or $U$-$V$ pair) was suggested:
  \beq\label{q04}
  \begin{array}{l}
  \displaystyle{
 U(z)=\sum\limits_{k=1}^3S_k\sigma_k\vf_k(z)\,,\qquad
 V(z)=\sum\limits_{k=1}^3S_k\sigma_k\frac{\vf_1(z)\vf_2(z)\vf_3(z)}{\vf_k(z)}+
 \sum\limits_{k=1}^3 W_k\sigma_k\vf_k(z)\,,
 }
 \end{array}
 \eq
 where $\vf_k(z)$ is the following set of elliptic functions on elliptic curve with moduli $\tau$
 (with ${\rm Im}(\tau)>0$):
 \beq\label{q05}
 \begin{array}{c}
  \displaystyle{
 %01:
 \vf_{1}(z)=e^{\pi\imath z}\phi(z,\frac{\tau}{2})\,,\quad
 %11
 \vf_{2}(z)=e^{\pi\imath z}\phi(z,\frac{1+\tau}{2})\,,\quad
 %10:
 \vf_{3}(z)=\phi(z,\frac{1}{2})\,.
 }
  \end{array}
 \eq
 The functions $\vf_k(z)$ are defined through the Kronecker elliptic function $\phi(z,u)$ and the first Jacobi theta-function $\vth(z)$:
\beq\label{q06}\begin{array}{c}
\displaystyle{
\phi(z, u) =\frac{\vartheta'(0) \vartheta (z + u)}{\vartheta (z) \vartheta (u)}\,,\quad
     \vartheta (z)=\vartheta (z|\tau) = -\sum_{k\in \mathbb{Z}} \exp \left( \pi \imath \tau (k + \frac{1}{2})^2 + 2\pi \imath (z + \frac{1}{2}) (k + \frac{1}{2}) \right)\,.
}\end{array}\eq
 The theta function is odd $\vth(-z)=-\vth(z)$ and has simple zero at $z=0$. Therefore, all $\vf_k(z)$
 have simple pole at $z=0$. Thus
$\res\limits_{z=0}U(z)=S$.
 Plugging the Lax pair (\ref{q04}) into equation (\ref{q03}) one gets two types of terms. The first type contains the first order pole at $z=0$, and the second type -- those terms with the second order pole at $z=0$. These two types
 terms vanish separately since (\ref{q03}) is assumed to be valid identically in $z$. In this way we get two equations. From vanishing of the terms with the first order pole one gets\footnote{To derive (\ref{q08}) one should use the identities $\vf^2_k(z)=\wp(z)-\wp(\om_k)$, where $\om_1=\tau/2$, $\om_2=(\tau+1)/2$, $\om_3=1/2$
 and $\wp(z)$ is the Weierstrass $\wp$-function.}:
 \beq\label{q08}
 \begin{array}{c}
  \displaystyle{
 \p_t S=\p_x T+[S,J(S)]\,,\qquad T=\sum\limits_{k=1}^3 T_k\sigma_k\,,
 }
  \end{array}
 \eq
 where $J_k=\wp(\om_k)$.
 From vanishing of the terms with the second order pole we obtain
 \beq\label{q09}
 \begin{array}{c}
  \displaystyle{
\p_x S=[S,T]\,.
 }
  \end{array}
 \eq
 The latter equation can be solved with respect to $T$. Since $S$ is a traceless $2\times 2$ matrix, it satisfies
 the characteristic equation
$
S^2=\lambda^2 1_2
$,
 where $1_2=\sigma_0$ is the identity $2\times 2$ matrix and $\lambda$ is an eigenvalue of $S$. Suppose
 $\p_x\lambda=0$. This means that the length of vector $\vec{S}$ is constant along the $x$ direction. Then,
 differentiating the characteristic equation with respect to $x$ yields
$
(\p_x S) S+S (\p_x S)=0
$.
 One can easily verify that equation (\ref{q09}) has the following solution:
$
 T=\frac{1}{4\la^2}[S,\p_x S]
 $.
 Plugging it into (\ref{q08}) we obtain the Landau-Lifshitz equation (\ref{q02}) with $c_1=1$, $c_2=1/(4\la^2)$.

 \paragraph{Purpose of the paper} is  to propose higher rank generalization of the above construction.
 For this purpose we represent the $U$-$V$ pair in terms of $R$-matrix data. Let us demonstrate it for the
 above example. The classical ${\rm gl}_2$  elliptic $r$-matrix has the form:
  \beq\label{q121}
 \begin{array}{c}
  \displaystyle{
 r_{12}(z)=\sigma_0\otimes \sigma_0\,E_1(z)+\sum\limits_{k=1}^3 \sigma_k\otimes\sigma_k\,\vf_k(z)\,,
 \qquad E_1(z)=\frac{\vth'(z)}{\vth(z)}\,.
 }
  \end{array}
 \eq
 Define
  \beq\label{q122}
 \begin{array}{c}
  \displaystyle{
 U(z)=\sigma_0 S_0 E_1(z)+\sum\limits_{k=1}^3 \sigma_k S_k\vf_k(z)
 }
  \end{array}
 \eq
 and notice that it differs from the one given in (\ref{q04}) by only the scalar term $1_2 S_0 E_1(z)$, which is not necessary
 since $S_0=\tr(S)/2$ has trivial dynamics due to (\ref{q02}).
 The expression (\ref{q122}) is rewritten in terms of $r$-matrix (\ref{q121}):
  \beq\label{q123}
 \begin{array}{c}
  \displaystyle{
  U(z)=U(S,z)=\frac{1}{2}\,\tr_2\Big(r_{12}(z)\stackrel{2}{S}\Big)\,,\qquad \stackrel{2}{S}=1_2\otimes S\,,\qquad
 S=\sum\limits_{k=0}^3 \sigma_k S_k\,.
 }
  \end{array}
 \eq
 Similarly, one can represent $V(z)$ in the form\footnote{For this purpose one should
 use the identities $\p_z\vf_i(z)=-\vf_j(z)\vf_k(z)$ valid for any set of distinct $i,j,k\in\{1,2,3\}$.}:
$
 V(z)=-\p_z U(S,z)+U(T,z)
$.
 Therefore, the existence of the Zakharov-Shabat representation for the Landau-Lifshitz equations is
 based on some properties of the classical $r$-matrix (\ref{q121}) only.
 Below we suggest a universal construction based on certain properties of $R$-matrices, such as associative Yang-Baxter equation. In this way we describe
 not only elliptic model but also a family of trigonometric and rational models, which correspond to $R$-matrices
 satisfying the set of required relations.

 %%%%%%%%%%%%%%%%%%%%%%%%%%%%%%%%%%%%%%%%%%%%%%%%%%%%%%%%%%%%%%%%%%%%%%%%%%%%%%%%%%%%%%%%
 %%%%%%%%%%%%%%%%%%%%%%%%%%%%%%%%%%%%%%%%%%%%%%%%%%%%%%%%%%%%%%%%%%%%%%%%%%%%%%%%%%%%%%%%
 \subsection*{\underline{Yang-Baxter equations}}

\paragraph{Quantum and classical $R$-matrices.} The Landau-Lifshitz equation is the classical continuous version of
 a certain quantum spin chain, which is described by the underlying quantum $R$-matrix. Quantum ${\rm GL}_N$ $R$-matrix in the fundamental representation is a matrix valued function\footnote{Here
 $\{E_{ij}\,;\, i,j=1...N\}$ is the standard basis in $N\times N$ matrices $\Mat$: $(E_{ij})_{ab}=\delta_{ia}\delta_{jb}.$}
  $
  R^\hbar_{12}(z)=\sum\limits_{ijkl=1}^N R_{ij,kl}(\hbar,z)E_{ij}\otimes E_{kl}\in\Mat^{\otimes 2}$
 of the Planck constant $\hbar$ and the spectral parameter $z$.
 By definition, $R$-matrix satisfies the quantum Yang-Baxter equation \cite{Baxter,TF}:
  \beq\label{q14}
  \begin{array}{l}
  \displaystyle{
 R^{\hbar}_{12}(z_1-z_2)  R^{\hbar}_{13}(z_1-z_3) R^{\hbar}_{23}(z_2-z_3) =
 R^{\hbar}_{23}(z_2-z_3)  R^{\hbar}_{13}(z_1-z_3)  R^{\hbar}_{12}(z_1-z_2) \,.
 }
 \end{array}
 \eq
In the above equation $R^{\hbar}_{ab}(z_a-z_b)$ is considered as an element of $\Mat^{\otimes 3}$.
It is nontrivial in two tensor components (the $a$-th and $b$-th), and the third one is filled by identity matrix. For example, $R^\hbar_{13}(z)=\sum\limits_{ijkl=1}^N R_{ij,kl}(\hbar,z)E_{ij}\otimes 1_N\otimes E_{kl}$.
We consider a class of $R$-matrices, which also satisfy the unitarity property
\begin{equation}\label{q15}
\begin{array}{c}
    R^{\hbar}_{12} (z) R^{\hbar}_{21} (-z)
    %= (\wp (z) - \wp (x))\, 1_N \otimes 1_N\stackrel{(\ref{diffsign})}{=}
    =N^2\phi(N\hbar,z)\phi(N\hbar,-z)\, 1_N \otimes 1_N
\end{array}
\end{equation}
and the skew-symmetry\footnote{The exchanging of indices from 12 to 21 means the exchanging of the tensor components.
 For example, $R^\hbar_{21}(z)=P_{12}R^\hbar_{12}(z)P_{12}=\sum\limits_{ijkl=1}^N R_{ij,kl}(\hbar,z)E_{kl}\otimes E_{ij}$, where $P_{12}$ is the permutation operator. For any pair of $N$-dimensional vectors $u,v$ the permutation operator acts as $P_{12}(u\otimes v)=v\otimes u$. For any pair of $N\times N$ matrices $A,B$: $P_{12}(A\otimes B)=(B\otimes A)P_{12}$. Explicitly, $P_{12}=\sum\limits_{i,j=1}^N E_{ij}\otimes E_{ji}$.}
  \beq\label{q16}
    \displaystyle{
  R^\hbar_{12}(z)=-R^{-\hbar}_{21}(-z)\,.
  }
  \eq
A solution of (\ref{q14}) is defined up to multiplication by a function, but this freedom is fixed in the
r.h.s. of (\ref{q15}), where the function $\phi$ is from (\ref{q06}). In the trigonometric or rational case this function turns into $\phi^{\rm trig}(\hbar,z)=\pi\cot(\pi z)+\pi\cot(\pi \hbar)$
and $\phi^{\rm rat}(\hbar,z)=\hbar^{-1}+z^{-1}$.
With the normalization (\ref{q15}) we also have the property that  $R_{12}^\hbar(z)$ has single simple pole in $z$
at $z=0$ and the residue is the permutation operator:
$
 \res\limits_{z=0}R^\hbar_{12}(z)=NP_{12}
$.
In the classical limit $\hbar\rightarrow 0$ we have the expansion
  \beq\label{q19}
  \begin{array}{l}
  \displaystyle{
R^\hbar_{12}(z)=\frac{1}{\hbar}\,1_N\otimes 1_N+r_{12}(z)+\hbar\,m_{12}(z)+O(\hbar^2)\,,
 }
 \end{array}
 \eq
where $r_{12}(z)$ is the classical $r$-matrix. Plugging (\ref{q19}) into (\ref{q14}) provides the classical Yang-Baxter equation
\beq\label{q20}
\begin{array}{c}
    \displaystyle{
    [r_{12}, r_{13}] + [r_{12}, r_{23}] + [r_{13}, r_{23}] = 0\,,
    \qquad r_{ij} = r_{ij}(z_i-z_j)\,.
    }
\end{array}\eq
The classical $r$-matrix has the following expansion near $z=0$:
  \beq\label{q21}
  \begin{array}{l}
  \displaystyle{
r_{12}(z) = \frac{1}{z}\, NP_{12} + r^{(0)}_{12} +  O(z)\,.
 }
 \end{array}
 \eq
 Due to (\ref{q16}) one can easily get the (skew)symmetry properties for the coefficients of expansions
 (\ref{q19}) and (\ref{q21}):
  \beq\label{q22}
  \begin{array}{l}
  \displaystyle{
r_{12}(z) = -r_{21}(-z)\,,\qquad
m_{12}(z) = m_{21}(-z)\,,\qquad
r^{(0)}_{12}=-r^{(0)}_{21}\,,\qquad
m_{12}(0) = m_{21}(0)
\,.
 }
 \end{array}
 \eq

\paragraph{Associative Yang-Baxter equation.} Besides the quantum Yang-Baxter equation (\ref{q14})  (which is
valid for any $R$-matrix) the $R$-matrices under consideration satisfy also the quadratic relation
  \beq\label{q24}
  \begin{array}{c}
    \displaystyle{
  R^{\hbar}_{12} R^{\eta}_{23} = R^{\eta}_{13} R^{\hbar-\eta}_{12} + R^{\eta-\hbar}_{23} R^{\hbar}_{13}\,,
   \qquad R^x_{ab} = R^x_{ab}(z_a-z_b)
 }
 \end{array}
 \eq
known as the associative Yang-Baxter equation \cite{FK}. Together with (\ref{q15}) and (\ref{q16}) equation
(\ref{q24}) provides the quantum Yang-Baxter equation (\ref{q14}), that is $R$-matrices under consideration
belong to a special subset of quantum $R$-matrices, see details in \cite{LOZ8,LOZR,LOZ15}. As was shown in \cite{LOZ2}, the equation (\ref{q24})  degenerates in a certain limiting
case into
$
[m_{13}(z_1)+m_{23}(z_2),r_{12}(z_1-z_2)]=[m_{12}(z_1-z_2)+m_{13}(z_1),r_{23}(z_2)]
$
 and in the limit $z_2\rightarrow z_1$ one gets
 \beq\label{q26}
 \begin{array}{c}
  \displaystyle{
 [m_{13}(z),r_{12}(z)]=[r_{12}(z),m_{23}(0)]-[\p_z m_{12}(z),NP_{23}]
  +[m_{12}(z),r_{23}^{(0)}]+[m_{13}(z),r_{23}^{(0)}]\,.
 }
 \end{array}
 \eq
Another important relation coming from (\ref{q24}) is as follows:
 \beq\label{q27}
  \begin{array}{c}
  \displaystyle{
  r_{12}(z)r_{13}(z\!+\!w)-r_{23}(w)r_{12}(z)+r_{13}(z\!+\!w)r_{23}(w)=m_{12}(z)+m_{23}(w)+m_{13}(z\!+\!w)\,.
 }
 \end{array}
 \eq
In particular, one can derive the classical Yang-Baxter equation (\ref{q20}) from (\ref{q27}) using the
symmetry properties (\ref{q22}). We are going to use degeneration of (\ref{q27}) in the limit $w\rightarrow 0$:
 \beq\label{q28}
  \begin{array}{c}
  \displaystyle{
  r_{12}(z)r_{13}(z)=r_{23}^{(0)}r_{12}(z)-r_{13}(z)r_{23}^{(0)}
  -N\p_z r_{13}(z)P_{23}+m_{12}(z)+m_{23}(0)+m_{13}(z)\,.
 }
 \end{array}
 \eq
%

 %%%%%%%%%%%%%%%%%%%%%%%%%%%%%%%%%%%%%%%%%%%%%%%%%%%%%%%%%%%%%%%%%%%%%%%%%%%%%%%%%%%%%%%%
 %%%%%%%%%%%%%%%%%%%%%%%%%%%%%%%%%%%%%%%%%%%%%%%%%%%%%%%%%%%%%%%%%%%%%%%%%%%%%%%%%%%%%%%%
\subsection*{\underline{Lax pairs and Zakharov-Shabat equation through $R$-matrices}}

\paragraph{Classical mechanics of integrable top.} We begin with the 0+1 example considered in \cite{LOZ2}. Introduce
 \beq\label{q29}
  \begin{array}{c}
  \displaystyle{
  L(S,z)=\frac{1}{N}\,\tr_2\Big(r_{12}(z)\stackrel{2}{S}\Big)\,,\qquad M(S,z)=\frac{1}{N}\,\tr_2\Big(m_{12}(z)\stackrel{2}{S}\Big)\,,\qquad
  \stackrel{2}{S}=1_N\otimes S
 }
 \end{array}
 \eq
where $S$ is an arbitrary $N\times N$ matrix, which matrix elements are dynamical
variables in the model of integrable top. Equations of motion have the form ${\dot S}=[S,J(S)]$, where
 \beq\label{q30}
  \begin{array}{c}
  \displaystyle{
  J(S)=\frac{1}{N}\,\tr_2\Big(m_{12}(0)\stackrel{2}{S}\Big)=M(S,0)\,.
 }
 \end{array}
 \eq
In order to show that the equations of motion are represented in the Lax for ${\dot L}(z)=[L(z),M(z)]$
one needs to compute $[L(z),M(z)]$. For this purpose we multiply both parts of (\ref{q26}) by
 $\stackrel{2}{S}\stackrel{3}{S}$
and take trace (of both parts) $\tr_{2,3}$ over the second and the third tensor components. Then
the last three commutators in the r.h.s. of (\ref{q26}) are cancelled out (see details in \cite{LOZ2}) and one gets
 \beq\label{q31}
  \begin{array}{c}
  \displaystyle{
 [L(S,z),M(S,z)]=L([S,J(S)],z)\,.
 }
 \end{array}
 \eq
In this way we conclude that the Lax equations hold true on the equations of motion. To prove converse statement
we mention that for any matrix $A$ the map $A\rightarrow L(A,z)$ is linear and $L(A,z)=0$ iff $A=0$.

\paragraph{Ansatz for $U$-$V$ pair.} In contrast to the above example in mechanics, in 1+1
case the set of matrices of dynamical variables $S$ is restricted by the following condition:
 \beq\label{q32}
  \begin{array}{c}
  \displaystyle{
 S^2=cS\,,
 }
 \end{array}
 \eq
where $c\in\mC$ is some constant. The condition (\ref{q32}) means that
the eigenvalues of the matrix $S$ are equal to either $0$ or $c$. In particular case, when $N-1$ eigenvalues coincide, we come to the matrix $S$ of rank 1. This case will be considered below separately.

Using the coefficient $r^{(0)}_{12}$ in the expansion (\ref{q21}) introduce the following linear map:
 \beq\label{q33}
  \begin{array}{c}
  \displaystyle{
 A\rightarrow E(A)=\frac{1}{N}\,\tr_2\Big(r^{(0)}_{12}\stackrel{2}{A}\Big)\,,\quad \stackrel{2}{A}=1_N\otimes A\,,\quad  A\in\Mat\,.
 }
 \end{array}
 \eq
It is important to mention that in the $N=2$ case explicit evaluation provides $r^{(0)}_{12}=0$, that is
$E(A)\neq 0$ for the higher rank cases ($N\geq 3$) only.
Define
 \beq\label{q34}
  \begin{array}{c}
  \displaystyle{
 U(z)=L(S,z)=\frac{1}{N}\,\tr_2\Big(r_{12}(z)\stackrel{2}{S}\Big)\,,\qquad
 V(z)=V_1(z)+V_2(z)\,,
 }
 \end{array}
 \eq
 \beq\label{q35}
  \begin{array}{c}
  \displaystyle{
 V_1(z)=-c\p_z L(S,z)+L(SE(S),z)+L(E(S)S,z)\,,\qquad
 V_2(z)=-cL(T,z)\,.
 }
 \end{array}
 \eq
\paragraph{Derivation of equations.} Multiplying both sides of the identity (\ref{q28}) by $\stackrel{2}{A}\stackrel{3}{B}$ and taking trace $\tr_{2,3}$ we get
 \beq\label{q36}
  \begin{array}{c}
  \displaystyle{
 L(A,z)L(B,z)=L(AE(B),z)+L(E(A)B,z)-\p_z L(AB,z)+
 }
 \\ \ \\
   \displaystyle{
 +\frac{\tr(B)}{N}\,M(A,z)
 +\frac{\tr(A)}{N}\,M(B,z)+\frac{1_N}{N}\,\tr_{23}\Big(m_{23}(0)\stackrel{2}{A}\stackrel{3}{B}\Big)\,,
 }
 \end{array}
 \eq
where $A,B\in\Mat$ are arbitrary. Plugging $A=B=S$ into (\ref{q36})
yields
%
% \beq\label{q37}
%  \begin{array}{c}
$$
  \displaystyle{
 L^2(S,z)=L(SE(S),z)+L(E(S)S,z)-\p_z L(S^2,z)
 +\frac{2\tr(S)}{N}\,M(S,z)
 +\frac{1_N}{N}\,\tr_{23}\Big(m_{23}(0)\stackrel{2}{S}\stackrel{3}{S}\Big)\,.
 }
 $$
% \end{array}
% \eq
%
Taking into account (\ref{q32}), this relation  provides
an alternative representation for $V_1(z)$ defined in (\ref{q35}):
 \beq\label{q38}
  \begin{array}{c}
  \displaystyle{
 V_1(z)=L^2(S,z)
 -2s_0\,M(S,z)
 -\frac{1_N}{N}\,\tr_{23}\Big(m_{23}(0)\stackrel{2}{S}\stackrel{3}{S}\Big)\,,\quad s_0=\frac{\tr(S)}{N}\,.
 }
 \end{array}
 \eq
Also, it is easy to see from (\ref{q36}) that
 \beq\label{q39}
  \begin{array}{c}
  \displaystyle{
 [L(S,z),L(T,z)]=-\p_zL([S,T],z)+L([S,E(T)],z)+L([E(S),T],z)\,.
 }
 \end{array}
 \eq
Plugging the ansatz for $U$-$V$ pair (\ref{q34})-(\ref{q35}) into the Zakharov-Shabat equation
(\ref{q03}) we get
 \beq\label{q40}
  \begin{array}{c}
  \displaystyle{
  \p_t S+c\p_x T-\p_x(SE(S)+E(S)S)=2s_0[S,J(S)]+c[S,E(T)]+c[E(S),T]
 }
 \end{array}
 \eq
 and
 \beq\label{q41}
  \begin{array}{c}
  \displaystyle{
  -\p_xS=[S,T]\,.
 }
 \end{array}
 \eq
The last equation comes as vanishing of $\p_zL(*,z)$ terms with the second order pole at $z=0$, while (\ref{q40}) appears from vanishing of $L(*,z)$ terms with the first order pole at $z=0$. Notice that for $\p_x V(z)$ term in the
Zakharov-Shabat equation we used $V_1(z)$ as given in (\ref{q35}), and in the term $[U(z),V(z)]$ we used
 $V_1(z)$  in the form (\ref{q38}). It is useful because from representation (\ref{q38}) we have $[U(z),V_1(z)]=-2s_0[L(S,z),M(S,z)]\stackrel{(\ref{q31})}{=}-2s_0 L([S,J(S)],z)$. In order to solve
 (\ref{q41}) with respect to $T$ we mention  that $(S-(c/2)1_N)^2=(c/2)^21_N$, that is similarly to $N=2$ case we obtain
 \beq\label{q42}
  \begin{array}{c}
  \displaystyle{
  T=-c^{-2}[S,\p_x S]\,.
 }
 \end{array}
 \eq
Finally, we substitute $T$ from (\ref{q42}) to (\ref{q40}):
 \beq\label{q43}
  \begin{array}{c}
  \displaystyle{
  \p_t S-\frac{1}{c}\,[S,\p^2_x S]-\p_x\Big(SE(S)\!+\!E(S)S\Big)\!=\!
  2s_0[S,J(S)]\!-\!\frac{1}{c}\,[S,E([S,\p_x S])]\!-\!\frac{1}{c}\,[E(S),[S,\p_x S]]\,,
 }
 \end{array}
 \eq
where $s_0=\tr(S)/N$. This is the higher rank Landau-Lifshitz equation. It is formulated through
the linear maps $J(S)$ (\ref{q30}), $E(A)$ (\ref{q33}), which arise in the $R$-matrix expansions
(\ref{q19}) and (\ref{q21}). The dynamical matrix $S$ is assumed to satisfy condition (\ref{q32}) with some
constant $c$.

 %%%%%%%%%%%%%%%%%%%%%%%%%%%%%%%%%%%%%%%%%%%%%%%%%%%%%%%%%%%%%%%%%%%%%%%%%%%%%%%%%%%%%%%%
 %%%%%%%%%%%%%%%%%%%%%%%%%%%%%%%%%%%%%%%%%%%%%%%%%%%%%%%%%%%%%%%%%%%%%%%%%%%%%%%%%%%%%%%%
\subsection*{\underline{Elliptic case}}
The most general is the elliptic case. It comes from the elliptic Baxter-Belavin $R$-matrix \cite{Baxter}
in the fundamental representation of ${\rm GL}_N$. Introduce the special matrix basis in $\Mat$:
$T_a=T_{a_1 a_2}=\exp\left(\frac{\pi\imath}{{ N}}\,a_1
 a_2\right)Q_1^{a_1}Q_2^{a_2}$, where $a=(a_1,a_2)\in\mZ_{ N}\times\mZ_{ N}$, and
$(Q_1)_{kl}=\delta_{kl}\exp(\frac{2\pi
 \imath}{{ N}}k)$, $(Q_2)_{kl}=\delta_{k-l+1=0\,{\hbox{\tiny{mod}}}\,
 { N}}$. In particular, $T_{(0,0)}=1_N$. The basis has the property $\tr(T_\al T_\be)=N\delta_{\al+\be,(0,0)}$.
 See details in \cite{Baxter} (see also Appendix in \cite{ZZ}). The quantum $R$-matrix is as follows:
 \beq\label{q44}
 \begin{array}{c}
  \displaystyle{
 R_{12}^\hbar(z)=\sum\limits_{a\in\,\mZ_{ N}\times\mZ_{ N}} T_a\otimes T_{-a}
 \exp (2\pi\imath\,\frac{a_2z}{N})\,\phi(z,\frac{a_1+a_2\tau}{N}+\hbar)\in{\rm Mat}(N,\mC)^{\otimes 2}\,.
 }
 \end{array}
 \eq
The classical limit (\ref{q19}) provides\footnote{On should use the expansion of $\phi(z,u)$ near $u=0$:
$\phi(z,u)=u^{-1}+E_1(z)+u\rho(z)+O(u^2)$.}
 \beq\label{q45}
 \begin{array}{c}
  \displaystyle{
 r_{12}(z)=E_1(z)1_N\otimes 1_N+\sum\limits_{a\neq(0,0)} T_a\otimes T_{-a}
 \exp (2\pi\imath\,\frac{a_2z}{N})\,\phi(z,\frac{a_1+a_2\tau}{N})
 }
 \end{array}
 \eq
and
 \beq\label{q46}
 \begin{array}{c}
  \displaystyle{
 m_{12}(z)=\rho(z)1_N\otimes 1_N+\sum\limits_{a\neq(0,0)} T_a\otimes T_{-a}
 \exp (2\pi\imath\,\frac{a_2z}{N})\,f(z,\frac{a_1+a_2\tau}{N})\,,
 }
 \end{array}
 \eq
 where $f(z,u)=\p_w\phi(z,w)|_{w=u}$ and $\rho(z)=(E_1^2(z)-\wp(z))/2$. Then one finds
 \beq\label{q47}
 \begin{array}{c}
  \displaystyle{
 r_{12}^{(0)}=\!\sum\limits_{a\neq(0,0)}\! T_a\otimes T_{-a}\Big(2\pi\imath\,\frac{a_2}{N}+E_1(\frac{a_1\!+\!a_2\tau}{N}) \Big)\,,
 \ \
  E(A)=\!\sum\limits_{a\neq(0,0)} \!T_a A_a\Big(2\pi\imath\,
  \frac{a_2}{N}+E_1(\frac{a_1\!+\!a_2\tau}{N}) \Big)\,,
 }
 \end{array}
 \eq
 \beq\label{q48}
 \begin{array}{c}
  \displaystyle{
 m_{12}(0)=\frac{\vth'''(0)}{3\vth'(0)}1_N\otimes 1_N-\sum\limits_{a\neq(0,0)} T_a\otimes T_{-a}
 E_2(\frac{a_1+a_2\tau}{N})\,,\quad
E_2(x)=-E_1'(z)=-\p_z^2\log\vth(z)\,,
 }
 \end{array}
 \eq
 \beq\label{q49}
 \begin{array}{c}
  \displaystyle{
 J(S)=\frac{\vth'''(0)}{3\vth'(0)}1_N S_{0,0}-\sum\limits_{a\neq(0,0)} T_a S_a
 E_2(\frac{a_1+a_2\tau}{N})=\frac{\vth'''(0)}{3\vth'(0)}\,S-\sum\limits_{a\neq(0,0)} T_a S_a
 \wp(\frac{a_1+a_2\tau}{N})\,.
 }
 \end{array}
 \eq
 In the last line we also used relation $E_2(z)=\wp(z)-\vth'''(0)/(3\vth'(0))$. Let us mention that in  $N=2$ case
 the classical $r$-matrix (\ref{q45}) turns into (\ref{q121}). Another important remark is that in $N=2$
 case $E(A)=0$ for any $A\in{\rm Mat}_2$ since $E_1(1/2)=0$ and $E_1(\tau/2)=E_1((\tau+1)/2)=-\pi\imath$.

 %%%%%%%%%%%%%%%%%%%%%%%%%%%%%%%%%%%%%%%%%%%%%%%%%%%%%%%%%%%%%%%%%%%%%%%%%%%%%%%%%%%%%%%%
 %%%%%%%%%%%%%%%%%%%%%%%%%%%%%%%%%%%%%%%%%%%%%%%%%%%%%%%%%%%%%%%%%%%%%%%%%%%%%%%%%%%%%%%%
\subsection*{\underline{The case ${\rm rank}(S)=1$}}

If ${\rm rank}(S)=1$ then the matrix $S$ is represented as $S=\xi\otimes\psi $, where $\xi$ and $\psi$ are
 $N$-dimensional vector column and row respectively. Then $S^2=\tr(S)S$, so that $c=\tr(S)=Ns_0$ in (\ref{q32})
 and (\ref{q43}).

\paragraph{Special property and simplification.} Up till now, we have not used one more property of the elliptic $R$-matrix (\ref{q44}). This is the Fourier symmetry (see e.g. \cite{Fourier}) $R_{12}^\hbar(z)P_{12}= R_{12}^{z/N}(N\hbar)$. Using the expansions
(\ref{q19}), (\ref{q21}) it provides the property $r_{12}^{(0)}=r_{12}^{(0)}P_{12}$. Together with the skew-symmetry
(\ref{q22}) it simplifies the Landau-Lifshitz equation (\ref{q43}) for the rank 1 case\footnote{For example, in this way one can deduce the properties $SE(S)=0$, $SE((\p_x S)S)=0$, $SE(S(\p_x S))=-c(\p_x S)E(S)$.}. Namely, the following identity holds:
 \beq\label{q50}
 \begin{array}{c}
  \displaystyle{
 c\p_x\Big(SE(S)+E(S)S\Big)+
  [E([S,\p_x S]),S]+[[S,\p_x S],E(S)]=2c[E(\p_x S),S]\,.
 }
 \end{array}
 \eq
 We will give its proof in our future papers. Using (\ref{q50}) the Landau-Lifshitz equation (\ref{q43})
 takes the form:
 \beq\label{q51}
  \begin{array}{c}
  \displaystyle{
  \p_t S=\frac{1}{c}\,[S,\p^2_x S]+\frac{2c}{N}\, [S,J(S)]-2[S,E(\p_x S)]\,.
 }
 \end{array}
 \eq
For $N=2$ case the last term in the r.h.s. vanishes and we come back to the original equation (\ref{q02}).

\paragraph{Hamiltonian description.} The equation (\ref{q51}) can be easily described in the Hamiltonian formalism.
The Poisson structure is given by
 \beq\label{q52}
  \begin{array}{c}
  \displaystyle{
 \{S_{ij}(x),S_{kl}(y)\}=(S_{kj}(x)\delta_{il}-S_{il}(x)\delta_{kj})\delta(x-y)
 \quad
 {\rm or}
 \quad
 \{\stackrel{1}{S}(x),\stackrel{2}{S}(y)\}=[P_{12},\stackrel{1}{S}(x)]\delta(x-y)\,.
 }
 \end{array}
 \eq
The equations of motion $\p_tS(x)=\{H,S(x)\}$ provide (\ref{q51}) for the following Hamiltonian:
 \beq\label{q53}
  \begin{array}{c}
  \displaystyle{
 H=\oint dy\Big( \frac{c}{N}\,\tr\Big(S\,J(S)\Big)-\frac{1}{2c}\,\tr\Big(\p_yS\,\p_yS\Big)+
 \tr\Big(\p_y S\,E(S)\Big)\Big)\,,\qquad S=S(y)\,.
 }
 \end{array}
 \eq
%
%where $S=S(y)$.

%\paragraph{Polynomial representation.}
%\setcounter{equation}{0}

\subsection*{\underline{Concluding remarks}}

We described a special class of higher rank generalizations of the Landau-Lifshitz equation.
Main idea was to use a set of $R$-matrix identities coming from the associative Yang-Baxter equation.
As a result we deduce equations of motion (\ref{q43}) and the Lax pairs (\ref{q34})-(\ref{q35}) with spectral parameter on elliptic curve (or its degeneration). In the particular case (${\rm rank}(S)=1$) the equation (\ref{q43}) is simplified to (\ref{q51}). The latter
admits quite simple Hamiltonian description (\ref{q52})-(\ref{q53}).

It should be mentioned that the study of general matrix valued Lax pairs for the Zakharov-Shabat equation on elliptic curves is known from \cite{Chered}. Notice that in the general case the Lax matrix has a set of poles (not a single one). Then the 1+1 Gaudin type models \cite{Z11} arise. We hope our construction can be generalized to this case as well.

Explicit formulae in the elliptic case are given in (\ref{q44})-(\ref{q49}). Details of trigonometric and rational degenerations will be given in our next paper.
The underlying finite-dimensional models (of type (\ref{q29})-(\ref{q31})) were studied in \cite{AASZ,KrZ,LOZ8,LOZR}. $N=2$ cases were considered in our previous paper \cite{AtZ}, where a gauge equivalence between the Landau-Lifshitz equations and the 1+1 Calogero-Moser field theories was described. In view of similar relation between the XYZ chain and the Ruijsenaars-Schneider model \cite{ZZ} we also expect to extend our construction to (semi)discrete equations.

The Landau-Lifshitz equation (\ref{q01}) has multicomponent generalization suggested by
I.Z. Golubchik, V.V. Sokolov \cite{GS} and T.V. Skrypnyk \cite{Skrypnyk}. We hope to clarify relation of our construction
to the results of \cite{GS} in our future works.

%%%%%%%%%%%%%%%%%%%%%%%%%%%%%%%%%%%%%%%%%%%%%%%%%%%%%%%%%%%%%%%%%%%%%%%%%%%%%%%%%%%%%%%%%%%%%%%%%%%%%%
%%%%%%%%%%%%%%%%%%%%%%%%%%%%%%%%%%%%%%%%%%%%%%%%%%%%%%%%%%%%%%%%%%%%%%%%%%%%%%%%%%%%%%%%%%%%%%%%%%%%%%

\subsection*{\underline{Acknowledgments}}
The work of A. Zotov is supported by the Russian Science Foundation under grant
21-41-09011.
%19-11-00062 and performed in Steklov Mathematical Institute of Russian Academy of Sciences.

%%%%%%%%%%%%%%%%%%%%%%%%%%%%%%%%%%%%%%%%%%%%%%%%%%%%%%%%%%%%%%%%%%%%%%%%%%%%%%%%%%%%%%%%%%%%%%%%%%%%%%
%%%%%%%%%%%%%%%%%%%%%%%%%%%%%%%%%%%%%%%%%%%%%%%%%%%%%%%%%%%%%%%%%%%%%%%%%%%%%%%%%%%%%%%%%%%%%%%%%%%%%%

\begin{small}

\end{small}


\begin{thebibliography}{99}
\addcontentsline{toc}{section}{References}

\bibitem{AASZ}  G. Aminov, S. Arthamonov, A. Smirnov, A. Zotov,
  %“Rational top and its classical r-matrix”,
  J. Phys. A: Math. Theor., 47:30 (2014), 305207;\\ arXiv: 1402.3189 [math-ph].
  \\
  A. Grekov, I. Sechin, A. Zotov, 	JHEP 10 (2019) 081; arXiv:1905.07820 [math-ph].

\bibitem{AtZ}
K. Atalikov, A. Zotov,
%{\em Field theory generalizations of two-body Calogero–Moser models in the form of Landau–Lifshitz equations},
 J. Geom. Phys. 164 (2021) 104161 , 14 pp., arXiv: 2010.14297 [math-ph].

\bibitem{Baxter}
R.J. Baxter,
% {\em Partition function of the eight-vertex lattice model},
Ann. Phys. 70 (1972) 193--228.
\\
A.A. Belavin,
%{\em Dynamical symmetry of integrable quantum systems},
Nucl. Phys. B, 180 (1981) 189--200.



%\bibitem{BD} A.A. Belavin, V.G. Drinfeld,
%{\em Solutions of the classical Yang–Baxter equation for simple Lie algebras},
%Funct. Anal. Appl., 16:3 (1982) 159--180.

\bibitem{Chered}
I.V. Cherednik,
% {\em Relativistically invariant quasiclassical limits of integrable two-dimensional quantum models},
Theoret. and Math. Phys.,  47:2 (1981) 422--425.
\\
I.V. Cherednik,
% {\em Elliptic curves and soliton matrix differential equations},
Journal of Soviet Mathematics, 38 (1987) 1989--2027.

\bibitem{GS} I.Z. Golubchik, V.V. Sokolov,
%{\em Multicomponent generalization of the hierarchy of the Landau-Lifshitz equation},
Theoret. and Math. Phys. 124 (2000) 909--917.

\bibitem{FT}
L.D. Faddeev, L.A. Takhtajan,
{\em Hamiltonian methods in the theory of solitons},
Springer-Verlag, (1987).

\bibitem{FK} S. Fomin, A.N. Kirillov,
%{\em Quadratic algebras, Dunkl elements, and Schubert calculus},
Advances in geometry; Prog. in Mathematics book series, 172
 (1999) 147--182.
\\
 A. Polishchuk, % {\em Classical Yang–Baxter equation and the $A^\infty$-constraint},
Advances in Mathematics 168:1 (2002)  56–-95.


\bibitem{KrZ} T. Krasnov, A. Zotov, Annales Henri Poincare, 20:8 (2019)
2671--2697;
% {\em Trigonometric integrable tops from solutions of associative Yang-Baxter equation},
 arXiv:1812.04209 [math-ph].


%\bibitem{Laksh} M. Lakshmanan,
% {\em The Fascinating World of Landau-Lifshitz-Gilbert Equation: An Overview}
%Phil. Trans. R. Soc. A.369 (2011) 1280--1300;
% arXiv:1101.1005 [nlin.PS]


\bibitem{LL} L.D. Landau, E.M. Lifshitz,
% {\em To the Theory of Dispersion of Magnetic Permeability of Ferromagnetic Solids. Collection of L.D. Landau Works}, Nauka %(1969), Vol. 1.
Phys. Zs. Sowjet., 8 (1935) 153--169.
% To the Theory of Magnetic Permeability Dispersion in Ferromagnetic Solids



\bibitem{LOZ8} A. Levin, M. Olshanetsky, A. Zotov,
%{\em Relativistic Classical Integrable Tops and Quantum R-matrices},
JHEP 07 (2014) 012,
arXiv:1405.7523 [hep-th].
\\
A. Levin, M. Olshanetsky, A. Zotov,
%{\em Classical integrable systems and soliton equations related to eleven-vertex R-matrix},
	Nuclear Physics B 887 (2014) 400--422;	arXiv:1406.2995 [math-ph].

\bibitem{LOZR} A. Levin, M. Olshanetsky, A. Zotov,
%{\em Planck Constant as
%Spectral Parameter in Integrable Systems and KZB Equations},
JHEP 10 (2014) 109; arXiv:1408.6246 [hep-th].

\bibitem{LOZ15} A. Levin, M. Olshanetsky, A. Zotov,
%{\em Quantum Baxter-Belavin R-matrices and multidimensional Lax pairs for Painleve VI},
Theoret. and Math. Phys. 184:1 (2015) 924--939;	arXiv:1501.07351.
% [math-ph].
\\
A. Levin, M. Olshanetsky, A. Zotov,
% “Yang–Baxter equations with two Planck constants”,
J. Phys. A: Math. Theor., 49:1 (2016) 014003;
%Exactly Solved Models and Beyond: a special issue in honour of R. J. Baxter's 75th birthday,
arXiv: 1507.02617.
% [math-ph].

\bibitem{LOZ2} A. Levin, M. Olshanetsky, A. Zotov,
% {\em Noncommutative extensions of elliptic integrable Euler-Arnold tops and Painleve VI equation},
J. Phys. A: Math. Theor. 49:39 (2016) 395202; arXiv:1603.06101.
% [math-ph].


\bibitem{Skl} E.K. Sklyanin,
%{\em On complete integrability of the Landau-Lifshitz equation},
Preprint LOMI, E-3-79. Leningrad (1979).

%E.K. Sklyanin,
% {\em Poisson structure of a periodic classical XYZ -chain},
%Journal of Soviet Mathematics, 46 (1989) 1664–-1683.


\bibitem{Skrypnyk} T.V. Skrypnyk,
%{\em Quasigraded lie algebras, Kostant–Adler scheme, and integrable hierarchies},
Theoret. and Math. Phys. 142(2): 275–288 (2005).
\\
T. Skrypnyk,
%{\em Infinite-dimensional Lie algebras, classical r-matrices, and Lax operators: Two approaches},
J. Math. Phys.,  54, 103507, (2013).



\bibitem{TF}
L. Takhtajan, L. Faddeev,
% The quantum method of the inverse problem and the Heisenberg XYZ model,
 Russ. Math. Surveys, 34:5 (1979) 11--68.
\\
E.K. Sklyanin,
% {\em Poisson structure of a periodic classical XYZ -chain},
Journal of Soviet Mathematics, 46 (1989) 1664–-1683.

%E.K. Sklyanin,
 %{\em Some algebraic structures connected with the Yang—Baxter equation},
%Funct. Anal. Appl. 16:4 (1982) 263--270.


\bibitem{ZZ} A. Zabrodin, A. Zotov,
%{\em Field analogue of the Ruijsenaars-Schneider model},
	arXiv:2107.01697 [math-ph].

\bibitem{ZaSh}
%V.E. Zakharov, A.B. Shabat,
% {\em Exact Theory of Two-dimensional Self-focusing and One-dimensional Self-modulation of Waves in Nonlinear Media},
% Soviet physics JETP 34:1 (1972) 62--69.
%\\
%V.E. Zakharov, A.B. Shabat,
%% {\em Interaction between solitons in a stable medium},
%Zh. Eksp. Teor. Fiz. 64, (1973) 1627--1639.
%\\
V.E. Zakharov, A.B. Shabat,
% {\em A scheme for integrating the nonlinear equations
%of mathematical physics by the method of the inverse scattering problem. I}
Funct. Anal. Appl., 8:3 (1974), 226–-235.
\\
V.E. Zakharov, A.B. Shabat,
% {\em Integration of nonlinear equations of mathematical physics by the method of inverse scattering. II}
Funct. Anal. Appl., 13:3 (1979), 166–-174.
\\
V.E. Zakharov (eds) ''What Is Integrability?'', Springer Series in Nonlinear Dynamics
%Springer Berlin, Heidelberg,
%Springer Series in Nonlinear Dynamics ''What Is Integrability ?''
(1991).

\bibitem{Z11}
A. Zotov,
%{\em 1+1 Gaudin Model},
SIGMA 7 (2011) 067; arXiv:1012.1072 [math-ph].
\\
A. Levin, M. Olshanetsky, A. Zotov,
%{\em 2d Integrable systems, 4d Chern-Simons theory and Affine Higgs bundles},
	arXiv:2202.10106 [hep-th].

\bibitem{Fourier}
A. Zotov,
%{\em Relativistic elliptic matrix tops and finite Fourier transformations},
	Modern Physics Letters A, 32 (2017) 1750169; 	arXiv:1706.05601 [math-ph].

\end{thebibliography}
\end{document}